# On the Relationship of Dichotomy of Mars and Occurrence of Dust Devils with Crustal Magnetic Fields

Shivam Saxena and Jayesh Pabari
Planetary Sciences Division (PSDN)
Physical Research Laboratory, Ahmedabad 380 009, India

**Abstract**

The dichotomy referred to as a partition or separation of a whole into two parts; and specifically, the dichotomy is very important feature of Mars between the Southern and Northern regions of Mars, and another thing that makes Mars very special that is the occurrence of Dust Devils on Mars. So, we studied and survey the dust devil's occurrence on Mars in different Martian Years on the whole Mars. We create a 2D map of Martian Surface and plot the coordinates where the dust devils are captured during their activity and also use those locations where they leave the tracks behind them after passing from those locations and those tracks commonly referred to as a Dust Devil's Tracks. So, we plot them in two different categories Direct Observations and Indirect Observations of Dust Devils and in the map, we've plotted the locations (coordinates) of DDs shows a variation in locations of occurrences with the Dichotomy the serpent like variation we observed and we find most of the dust devil are occurred on the Dichotomy and the nearby regions of it which follows the serpent like trajectory of dichotomy of Mars and another observation shows that these locations lie on the remanent magnetic fields zones of mars which referred to as crustal magnetic fields of Mars this previously unknown relationship between crustal magnetic fields, dichotomy of mars and occurrence of dust devils is being examined here.

**Keywords:** Dichotomy; Dust Devils (DDs); Martian Years; Topography; Mars Crustal Magnetism.

## 1. Introduction

The dichotomy is referred to as a separation between two things or we can say that it's a partition of a whole thing into two separate parts which are different to each-other. So, the Hemispheric Dichotomy of Mars is a very interesting feature which separates the Northern areas of Mars by the Southern areas of Mars and this Dichotomy shows almost clear separation between heavily cratered region of Southern Hemisphere with the relatively smooth plain areas of Northern Hemisphere of Mars. (*Thomas et al. 2007*)

And this asymmetry which is described by (*Carr, 1981*) as North-South Asymmetry. And this asymmetry was very clear from the global image mosaic of mars which is by Mariner 9 (*Mutch et al. 1976*). And some recent High Contrast coloring of base Map of Global Topography of Mars is created using data provided by MOLA (Mars Orbiter Laser Altimeter) on NASA's Mars Global Surveyor which shows the variation and extent of the contrast between the Southern & Northern Hemisphere of Mars and it's a very fundamental crustal feature of Mars.

The Dust Devils which make Mars more important for exploration. The dust devils are the atmospheric convective vortices which can lift the dust from the surface and these vortices are resulted from the daytime surface heating and turbulence in the planetary boundary layer and as the air heated up and start to rising in upward direction and the vorticity becomes more vertical and start to intensifies and the air and whirling around a low-pressure region that develop in the vortex core (*Balme & Greeley, 2006; Kanak, 2005; Spiga et al., 2016; Toigo et al., 2003*). (*Newman et al. 2019*) and the study of Martian dust devils provides valuable insights into atmospheric processes on the planet. By analyzing their characteristics, such as size, duration, and frequency, scientists can better understand the dynamics of the Martian atmosphere, including its convective activity and the interaction between the surface and the atmosphere. Moreover, the lifting of dust particles by dust devils on Mars has implications for the planet's climate and the redistribution of dust on a global scale (*Greeley et al. 1997*) and when these dust devils are moving on the surface they leave behind some streaks or tracks which are referred to as Dust Devil Tracks (DDTs) and these albedo tracks or markings characterized by dark and bright tracks of dust devils and even when the dust vortices don't lift the enough dust from the surface or they do not carry enough dust particles to make them visible as dust devils do, these vortices can leave the passages on the surface which may lead to the tracks which can referred as same as DDTs for the understanding and simplicity. (*Perrin et al. 2020*)

The occurrence of dust devils on mars is very frequent and the dust devil's activity has been captured during the multiple NASA's missions. The NASA's Rover (Curiosity, Opportunity, Perseverance, Spirit), Landers (InSight) and Orbiters (Mars Reconnaissance Orbiter) and we try to get the coordinates of those locations where these dust devil's activity has been captured during the different mission sols.

And there is something which is very related to the occurrence of dust devils and the dichotomy of mars and locations where these dust devil's activity has been detected or captured are on the zones of remanent magnetic fields which may referred to as Mars Crustal magnetism or magnetic fields.

Mars currently has no Global magnetic field of internal origin but must have had one in the past (*Connerney et al. 2005*). And the Mars Global Surveyor (MGS) mission (1996-2006) led to many discoveries and new insights of Mars and one of them was enigmatic crustal magnetic field (*Acuña et al., 1998*) because before the mission of Mars Global Surveyor it was unknown that does Mars have its' own internal magnetic field or not. Like in some researches which is based on sparse measurement available described a weak planetary field (*Dolginov & Zhuzgov, 1991; Moehlmann et al., 1991; Slavin et al.,*

*1991*) and in some of the researches they argued for localized magnetic anomalies (*Moehlmann, 1992*).

But the Mars has its own feature of magnetic-field we can say it is a signature field of Mars and the maps and models which has been created by using MGS measurements which shows the crustal magnetic field zones over the terrain of mars and most of the highly magnetized terrains are found in the southern zone of Mars, and in the northern zone of Mars the magnetic field is lacking. (*Benoit Langlais et al. 2019*) So, Crustal magnetism is can be useful to understand the dust devils and their occurrences on the Mars.

In our observational survey and by our data analysis we tried to show that the three most important things or features of Mars we can say that are: Dust Devils, Dichotomy of Mars & Crustal Magnetic fields; which can play an important role to understand the dust devil's formation and occurrence on the certain locations on Mars and we manage to observe that most of the dust devils are occurred on the dichotomy of Mars and nearby regions of it and these are the zones of crustal magnetic fields which follows the dichotomy boundary pattern.

## 2. Methodology

To observe the relation between the dichotomy of Mars and the occurrence of dust devils on mars with the patches of crustal magnetic field on mars, we first try to create a dataset of the dust devil occurrences on Mars which is based on the observations of dust devils by NASA's rovers, landers, and orbiters.

The data of those mission which captured the images of mars surface and the Martian environment and in some of the images they also captured some dust devil's activity on Mars as well. We categorized those images as the dust devil activity on Mars and, by analyzing the multiple images we have taken those images in our dataset making.

But there is problem occurred during the acquisition of those images because many of the images doesn't have the locations information like where the dust devils occurred or captured during its activity on Mars. So, we try to extract those coordinates by did some analysis like we use the mission sols of the NASA mission and try to reach out at those locations by using the MARS TREK to pin point that locations with the precision of $\pm 0.5^{\circ}$ degrees radius region and, based on the availability of mission data we try to get coordinates as precise as we can.

To do that we use NASA's mars Rovers & Landers (Spirit, Curiosity, Opportunity, Perseverance) & HiRISE (*Hepburn et al. 2019*) images data and information to get the idea that where the dust devils are captured and traced during the mission sols. But almost every mission doesn't provide the coordinates of these dust devils like where they are captured on mars. So, by studying the NASA's articles, news, research papers and mission reports we try to derive or get to those coordinates.

When we say direct encounter of dust devils it means that the dust devils are captured during their activity by NASA's Rovers, Landers, or Orbiter and when we say Indirect encounter of dust devils it means we are observing the tracks of dust devils that they leave behind them by passing on the Martian surface and to do so, we categorized the observations into two categories:

1) Direct Observations or Direct Interactions with Dust Devils during NASA's missions.

2) Indirect Observations or Indirect Interactions with Dust Devils during the NASA's mission.

So, in these two categories first we acquire the relevant data we need to understand where exactly or maybe the encounters with dust devils at the different locations on Mars happened.

In these datasets we use different fields to categorize each event mentioned in the table below.

- **Direct Observations:**

In this category of observations, we go through the direct encounters of dust devils with NASA's Rovers, Landers & Orbiter on Mars. We go through the process to derive or get those coordinates which we describe further.

- **Indirect Observations:**

In this category of observation, we go through the indirect encounters of dust devils with NASA's HiRISE (High Resolution Science Imaging Experiment) which is the very powerful camera onboard on NASA's MRO (Mars Reconnaissance Orbiter) on Mars. we observed multiple HiRISE images to observe the tracks of dust devils and used the DDTs as data because these tracks are suggested that in the previous Martian Years there is dust devil activity on Mars. So, we can also take it as dust devil activity.

## 3. Making of Datasets

In making of the dataset, we go through some processes to get to the coordinates of occurring of dust devils on Mars given below.

We acquired the information about which rover, lander or orbiter capture the active dust devil or the dust devil tracks on the Martian Surface. Then we trace the whole mission track of that Rover or Lander to reach at that mission sol where the Rover or Lander captured the Dust Devil activity during its mission sol.

Then we use Mars Trek to point out that location and reach at the coordinates and the coordinates with precision of $\pm 0.5^{\circ}$ degrees.

And sometimes we are not able find the location because the location is not available so, we then we match the surface images which has taken by Rover or Lander and Match that surface Image by using Mars Trek[1] by finding that region on Mars and after matching the surface with the original image we're able to get the coordinates.

And for the Indirect observations as we already mentioned we use multiple HiRISE images to get to those locations. And after the data analysis, we take those images as data where we able to see the dust devil's tracks on the Martian surface. So, this is the way we get the coordinates of the Dust Devils.

### 3.1 Dataset of Direct Observations

The fields that we have taken for data acquisition are A-Date (Acquisition Date), Longitudes, Latitudes, Ls (Solar Longitude), MY (Martian Year), M-Sol (Mission Sol), O-Sol, Mission. In Table-1 there is a snippet of Direct Dust Devil's observation Dataset during their activity according to the fields we have taken. For our study we have taken

[1] It is a 3D Terrain Model of Martian Surface which is developed by NASA.

16 dust devils' activity as direct observations which is based on NASA's different missions which are mentioned in the Table-1.

**Acquisition Date:** This field shows when the data has been captured by NASA's Rovers, Landers, or Orbiter on Mars.

**Longitudes & Latitudes:** So, this is the very important data that we almost derive from the various sources using various applications to derive or get the coordinates of dust devil activity on Mars with the precision of ± 0.5° degrees and all the longitude values we taken as (E) eastward in direction.

**Ls:** So, it's called Solar Longitude or The Areocentric Longitude (Ls) is the longitude of the Sun as viewed from the center of Mars. It has a value from $0^0$ to $360^0$ and the range of the degrees represents the seasons on Mars.

**MY:** It's called Martian Year So, basically the 1 Martian year comprises 668.59 Martian days or sols (equivalent to 687 Earth days or 1.88 Tropical Earth years).

**M-Sol:** So, it means Mission-Sol or M-Sol. So, basically when NASA rover and landers are start to working on Mars after landing. So, one complete day on mars after landing of rover or lander called 1 Mission Sol.

**Observed-Sol:** So, in this field we put the approximate values of M-Sols, because of the unviability of exact Sol values for some missions and approximate the coordinate values for the encounters of Dust devils at a particular Mission-Sol.

**Mission:** In this field we put the mission's name of NASA's rover, lander, or orbiter. Which is working at that time and dust devil activity captured by that lander, Rover. or orbiter on Mars.

In the Table-1 all these fields and their related data associated with them which is going to be useful for the plotting out dust devils' locations on Mars and our further analysis.

*Table 1 Dataset for Direct Observations of Dust Devils during their activity on Mars*

| # | A-Date | Longitude | Latitude | Ls | MY | M-Sol | O-Sol | Mission |
|---|---|---|---|---|---|---|---|---|
| **1** | 10-Mar-2005 | 175.5201 | -14.5904 | 172.9 | 27 | 421 | Nearby Region | Spirit |
| **2** | 15-Apr-2005 | 175.5239 | -14.5903 | 193.3 | 27 | 456 | Nearby Region | Spirit |
| **3** | 18-Apr-2005 | 175.5252 | -14.5894 | 195.1 | 27 | 459 | Nearby Region | Spirit |
| **4** | 25-Jun-2005 | 175.5278 | -14.6049 | 236.8 | 27 | 525 | Nearby Region | Spirit |
| **5** | 02-Jul-2005 | 175.5254 | -14.5943 | 241.2 | 27 | 532 | Nearby Region | Spirit |
| **6** | 07-Jul-2005 | 175.5327 | -14.5926 | 244.4 | 27 | 537 | Nearby Region | Spirit |
| **7** | 13-Jul-2005 | 175.5259 | -14.6035 | 248.2 | 27 | 543 | Nearby Region | Spirit |
| **8** | 29-Jul-2005 | 175.5289 | -14.5922 | 258.4 | 27 | 559 | Nearby Region | Spirit |
| **9** | 26-Feb-2007 | 175.5320 | -14.6074 | 190.3 | 28 | 1120 | Nearby Region | Spirit |
| **10** | 09-Aug-2020 | 137.3911 | -4.7295 | 254.5 | 35 | 2847 | 2829 | Curiosity |
| **11** | 01-Feb-2017 | 137.3578 | -4.7108 | 309.3 | 33 | 1597 | 1596 | Curiosity |
| **12** | 18-Feb-2017 | 137.3585 | -4.7112 | 319.1 | 33 | 1613 | 1612 | Curiosity |
| **13** | 31-Mar-2016 | -5.3529 | -2.3158 | 130.1 | 33 | 4332 | 4330 to 4334 | Opportunity |
| **14** | 27-Sep-2021 | 77.4421 | 18.4338 | 104.6 | 36 | 215 | 2829 | Perseverance |
| **15** | 20-Jul-2021 | 77.4520 | 18.4283 | 74.1 | 36 | 148 | 1596 | Perseverance |
| **16** | 01-Feb-2019 | 135.8112 | 4.4755 | 333.6 | 34 | 65 | 1612 | Insight |

*Nearby Region means the area we have taken for coordinate plotting is around the mission sols because the coordinates are overlapping due to the very close coordinate values.

### 3.2 Dataset of Indirect Observations

The fields that we have taken for data acquisition in Indirect observations of Dust devils by using the HiRISE (High Resolution Imaging Science Experiment) images and the fields that are used for the acquisition of data are A-Date (Acquisition Date), LMT (Local Mars Time), Longitudes, Latitudes, Ls (Solar Longitude), MY (Martian Year), Image File Name and the definition and meaning of the fields is already described above.

In this dataset we have taken multiple images which can shows the dust devil tracks So, Dust devils that leave dark- or light-toned tracks are common on Mars. Dust devil tracks are surface features with mostly sub-annual lifetimes. The Dust devil's tracks have different sizes, Dust Devil's widths can range between ~1 m and ~1 km, depending on the diameter of dust devil that created the track, and the Dust Devil's lengths range from a few tens of meters to several kilometers, limited by the duration and horizontal ground speed of dust devils (*Dennis Reiss, Lori Fenton Lynn, Neakrase et al. 2016*).

In making of this dataset, we have taken 64 images from more than 350+ of images after analysis and observation and in these images the we found dust devil tracks and, in some images, the active dust devils are captured. So, those images we treated as direct observations of dust devils and put them into the direct observation category while creating the map.

*Table 2 Dataset for Indirect Observations of Dust Devils after their activity on Mars*

| # | A-Date | LMT | Longitude | Latitude | Ls | MY | Image File Name |
|---|---|---|---|---|---|---|---|
| **1** | 23-Dec-2008 | 15:48 | 29.049 | 14.625 | 178.5 | 29 | ESP_011289_1950 |
| **2** | 08-Mar-2009 | 15:52 | 26.822 | -55.076 | 223.1 | 29 | ESP_012252_1245 |
| **3** | 24-May-2009 | 15:09 | 168.260 | -52.562 | 272.2 | 29 | ESP_013249_1270 |
| **4** | 01-Jun-2009 | 15:05 | 346.238 | -61.083 | 277 | 29 | ESP_013348_1185 |
| **5** | 13-Jun-2009 | 14:39 | 175.541 | -14.633 | 284.3 | 29 | ESP_013499_1650 |
| **6** | 16-Jun-2009 | 15:07 | 11.428 | -68.603 | 286.5 | 29 | ESP_013545_1110 |
| **7** | 02-Jul-2009 | 14:50 | 145.020 | -68.531 | 296.3 | 29 | ESP_013751_1115 |
| **8** | 05-Jul-2009 | 14:40 | 293.100 | -49.804 | 297.9 | 29 | ESP_013785_1300 |
| **.** | . | . | . | . | . | . | . |
| **.** | . | . | . | . | . | . | . |
| **61** | 30-Aug-2008 | 15:24 | 199.486 | 32.849 | 120 | 29 | PSP_009819_2130 |
| **62** | 18-Sep-2008 | 15:28 | 182.748 | 23.914 | 128.8 | 29 | PSP_010057_2040 |
| **63** | 30-Sep-2008 | 15:42 | 33.237 | -37.566 | 135.0 | 29 | PSP_010221_1420 |
| **64** | 05-Nov-2008 | 15:37 | 120.148 | 40.322 | 152.9 | 29 | PSP_010679_2205 |

## 4. Mapping of Coordinates

After creating the datasets for both direct and indirect observations of dust devils we created a 2D map of Martian Surface from -180° to +180° degrees Longitude & -90° to +90° degrees latitude and put the coordinates to locate each dust devil based on our dataset. So, we created two separate maps, one based on direct observations of dust devils on Mars and other one is based on indirect observations on Mars and the meaning of terms is already described in above sections.

So, Figure-1 shows the map of active dust devils on Martian surface captured during different Missions in multiple Martian Years, each point on the map shows the location of dust devil encounter by NASA's Rover, Landers, or Orbiter which we already mentioned in the datasets above so, here in the Figure-1 each point represented by a color code which shows the Martian Year for that particular location and sometimes there are multiple encounters of dust devils in the same Martian Year repeatedly so, on some locations there is an overlapping of coordinates.

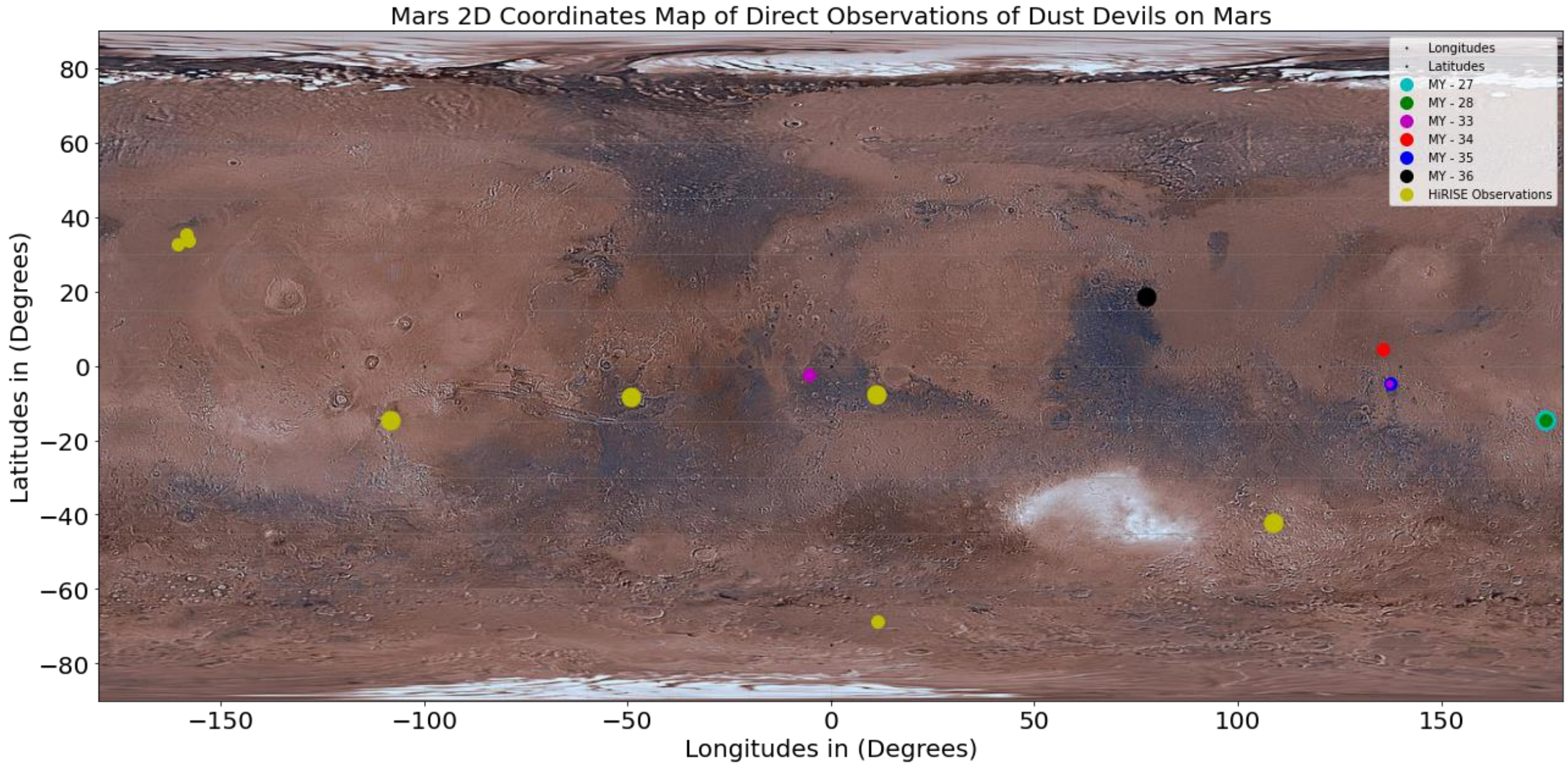

*Figure 1 Dust Devil's Direct Observations plotted locations on Mars Viking Colorized Global Mosaic, NASA AMES*

Similarly in Figure-2 each point represents the location of dust devil tracks on Mars. So, both the maps show the dust devil's activity on Mars in different Martian Years and the repetition of the dust devils in the same Martian Year like we can it the frequency of occurring of dust devils on some location in a particular Martian Year.

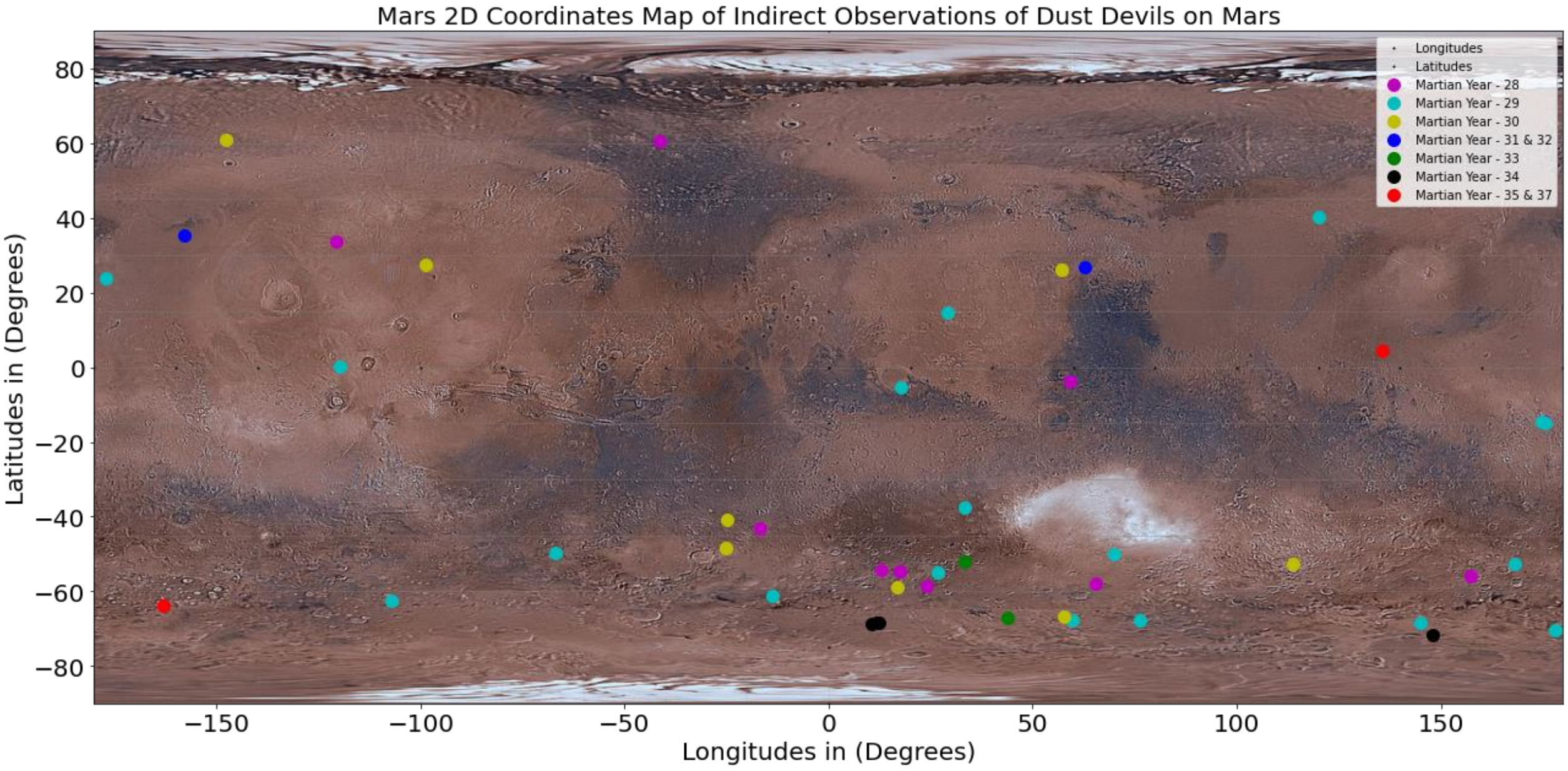

*Figure 2 Dust Devil's Indirect Observations plotted locations on Mars Viking Colorized Global Mosaic, NASA AMES*

So, in Figure-1 & Figure-2 which is plotted out on the Mars Viking Colorized Global Mosaic (https://astrogeology.usgs.gov) to give an idea about the position of dust devils on Mars with the Martian surface to get the better understanding of it but it is hard to see the pattern and the variation of the dust devil occurrences on the Martian surface. So, later we put these coordinates on the global topography of Mars based on data from the Mars Orbiter Laser Altimeter (MOLA) on NASA's Mars Global Surveyor.

## 5. Observational Survey of the Dust Devils occurrences

So, in our survey we have taken 80 Dust Devils activity (both direct & indirect observations) occurrences as we plotted on the maps so, when we combine both the maps and observe the variation in the dust devil occurrences on Mars is mostly followed by the Dichotomy of Mars. The dust devil's location as observed by superimposing the maps together as shown in Figure-4 with the map of the global topography of Mars. Coloring of the base map indicates relative elevations which is based on data of the Mars Orbiter Laser Altimeter (MOLA) on NASA's Mars Global Surveyor (MGS). So, Whites and browns indicate the highest elevations (+12 to +8 km); followed by pinks and reds (+8 to +3 km); yellow is 0 km; greens and blues are lower elevations (down to −8 km).

So, in the Figure-3 which shows the dichotomy of Mars which is clearly visible in this high contrast image. We projected dust devil's locations both Direct & Indirect observations and survey it and what we found is most of the dust devils are occurred or found during their activity or after activity on this dichotomy of Mars and the nearby region of that dichotomic boundary which is followed by the dust devils like the irregular path of the dichotomy.

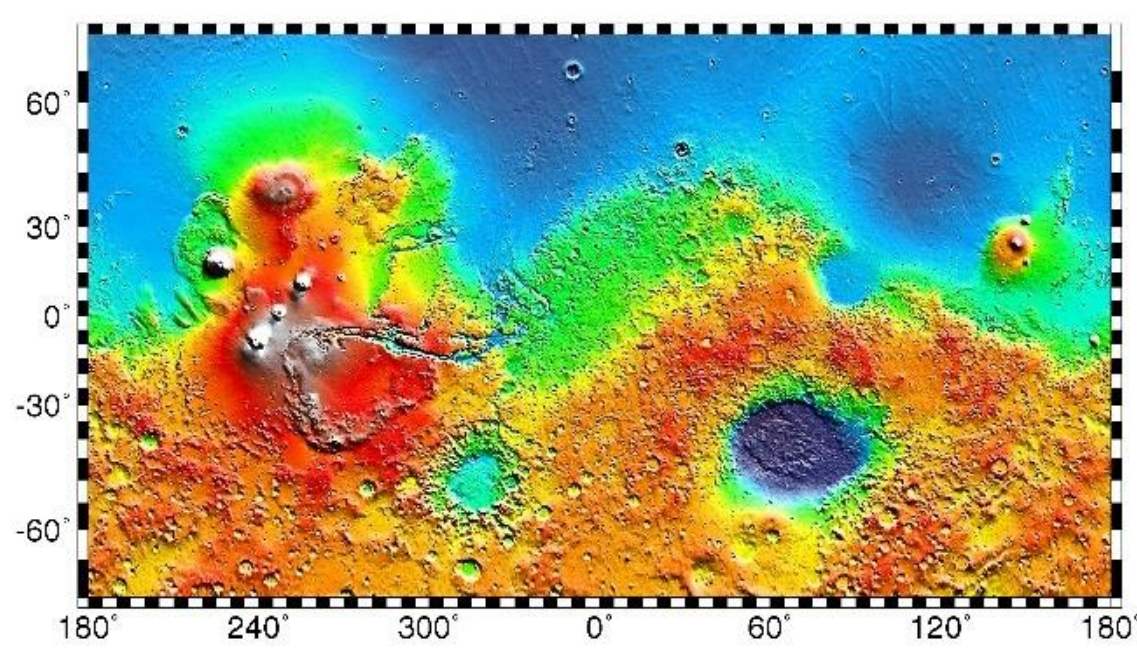

*Figure 3 Map of Mar's global topography, NASA/JPL/GSFC*

In Figure-4 where we superimpose all the 80 dust devils' activity with a single-color code for better visibleness and observations and to see the variation of these dust devils with the dichotomic boundary and the nearby regions of it.

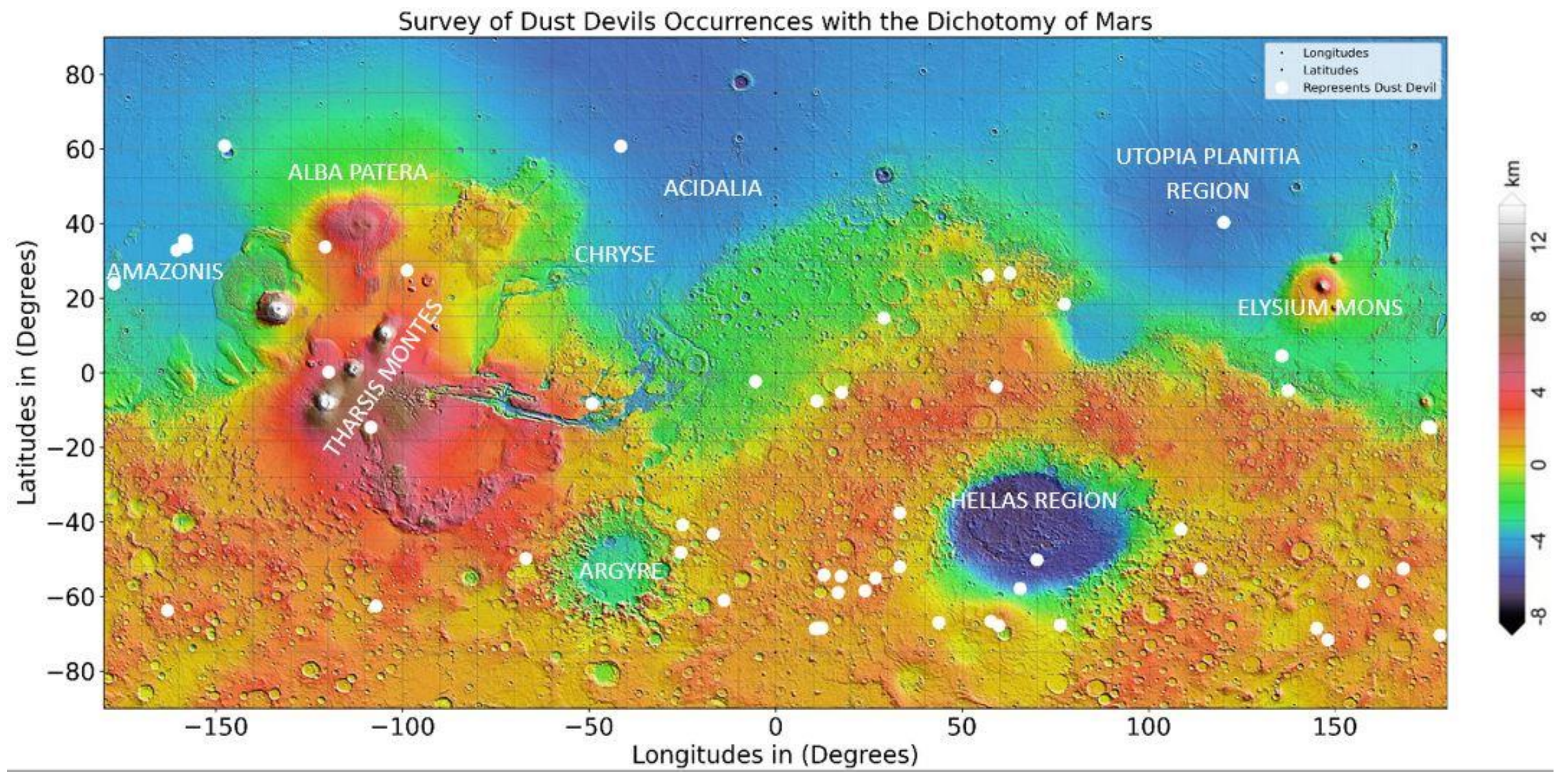

*Figure 4 Projection of Dust Devil's Locations (Coordinates) on the Mars topography (MOLA dataset) HiRISE Map, NASA/JPL/GSFC*

## 5.1 Survey of Dust Devils on Dichotomy of Mars & its nearby regions

In the Figure-4 the superimposed map of dust devils on the topographical map of Mars which is created using MOLA observations (*Courtesy: NASA/JPL-Caltech*). So, in this observational survey what we observe is that the locations of dust devils are such that they are following the dichotomy of Mars and the nearby regions of it so, we try to connect the dots which represents the locations of Dust Devils occurrences and try to show the pattern that the dust devils are following in the Figure-5.

As we can see in Figure-5 that the pattern followed by dust devils as they occurred on Mars, they mostly followed the dichotomy of Mars and the Upper-Shaded band in Figure-5 which shows the dust devils on the boundary or dichotomy of Mars and the Lower-Shaded band also mostly follows that serpent like dichotomic trajectory. Our data analysis shows that approximate ~70% ± 5% dust devils followed the dichotomy of Mars and mostly on the yellow and light red regions of the Mars Topography Map which is quite large number and when we observe the color bar which shows the relative elevation and depth of the surface in kilometers with dust devils' position on map what we observed is most of the dust devils are occurred in -4 to +4 km range of relative elevation and depth zones and quite frequent in the yellow regions which represents 0 km relative elevation. So, approximately ~85% ± 3% dust devils are observed and surveyed in the regions of -4 to + 4 km range of relative elevation and depth. So, a very good percentage of the Dust devils are found on the dichotomy of Mars and nearby regions of it and the dust devils which are not properly on the dichotomy of Mars follows the trajectory of the dichotomy boundary and another factor which follows an important role in the locations of dust devils which is the crustal magnetic fields of Mars.

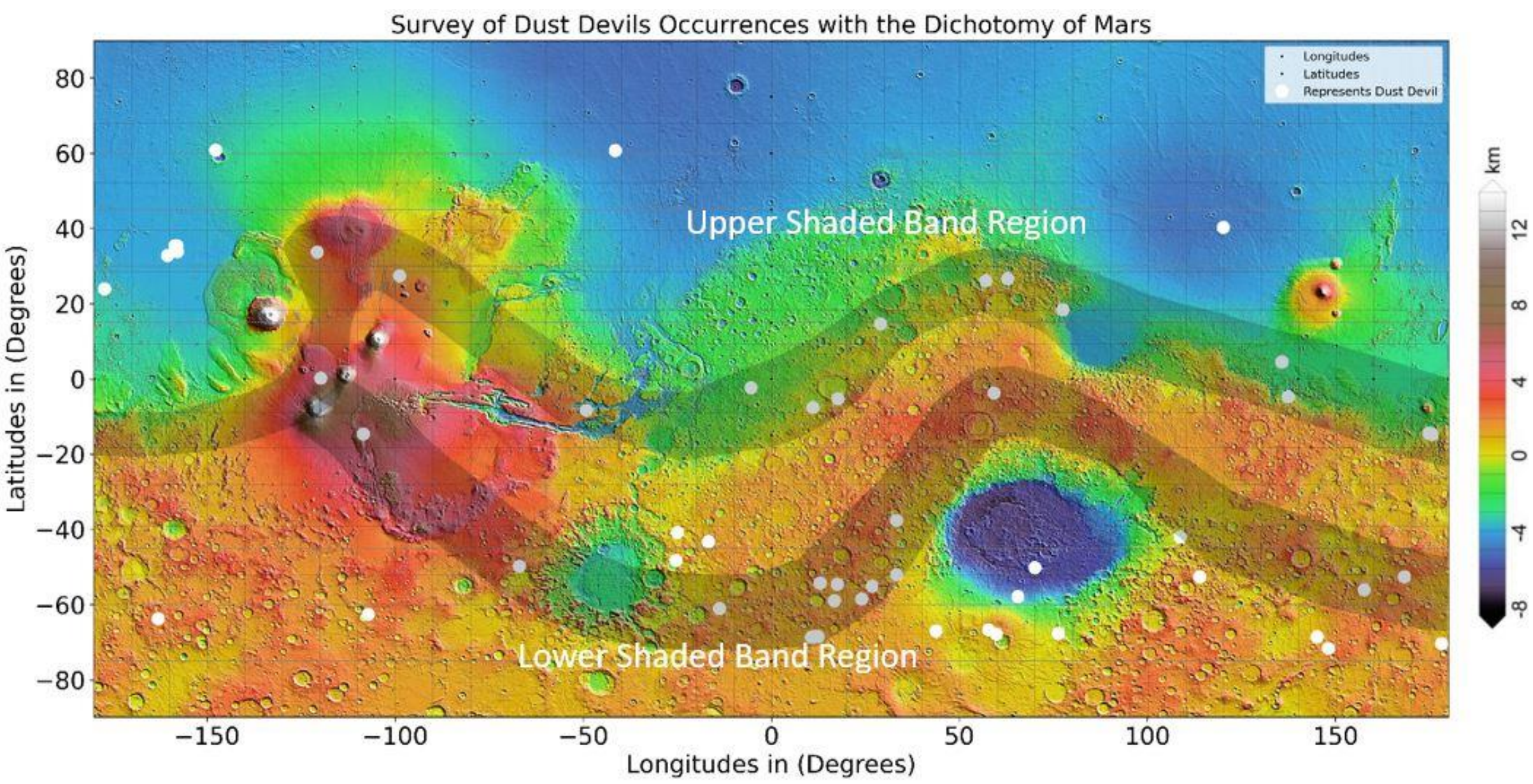

*Figure 5 Connecting the Dots: Locations (Coordinates) of Dust Devils on the Mars topography (MOLA dataset) HiRISE Map, NASA/JPL/GSFC*

## 5.2 Survey of Dust Devils on the Crustal Magnetic Field Zones of Mars

The crustal magnetic field is the remanent magnetic fields that Mars has in some of the regions and most of the highly magnetized terrains are found in the southern zone of Mars, and in the northern zones of Mars the magnetic field is lacking. (*Benoit Langlais et al. 2019*). In the Figure-6, (Br) is the radial field component of the magnetic field, whereas the median value of this quantity (ΔBr) is an estimate of the change in the radial magnetic field along track of the observation by MGS. And this map is superposed on the Mars Orbiter Laser Altimeter (MOLA) shaded topography map (*David E. Smith et al. 1999*) and which appears where ΔBr falls below a threshold value, which is ±0.3 nT per degree of latitude ($5 \times 10^{-3}$ nT/km) which traversed by MGS (Mars Global Surveyor). And in this map ΔBr component is of the vector field (Br, Bθ, Bφ) and may be obtained from the gradient of a scalar potential function, V, which is represented by a spherical harmonic expansion and can be written as Equation 1:

*Equation 1*

$$V = a \sum_{n=1}^{\infty} \left(\frac{a}{r}\right)^{n+1} \sum_{m=0}^{n} P_n^m(\cos\theta)[g_n^m \cos(m\varphi) + h_n^m \sin(m\varphi)]$$

The value of ΔBr is closely associated with the value of Bθ and the value of ΔBr is varied with the latitude in the map (*J. E. P. Connerney et al. 2005*). So, when we projected the dust devils on the Mars's crustal magnetic fields map which is the remanent magnetic fields zones on Martian surface in Figure-6. We observed from this survey that a huge amount of the dust devils approximates ~85% ± 2% are occurred or captured on the crustal magnetic field regions of Mars.

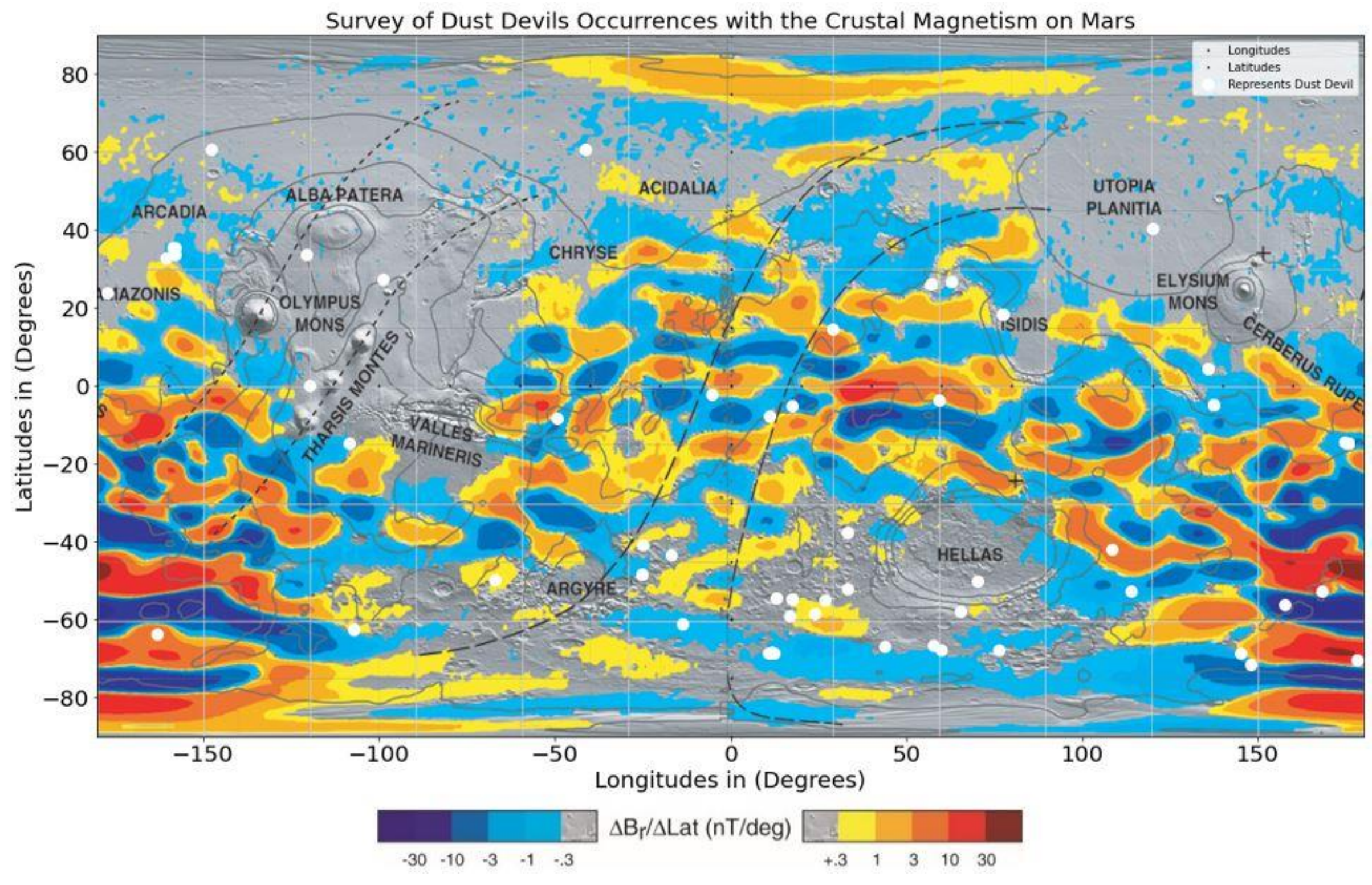

*Figure 6 Projection of Dust Devils on Mars Crustal Magnetic Fields Map MGS, NASA*

By our observational survey, it is shown that there is a relationship between the occurrence of dust devils with the dichotomy conditions and with the crustal magnetic field zones of Mars because the most probability of occurrence of dust devils' locations can be found in these regions of Mars where the crustal magnetic fields zones are there and these magnetic fields rich regions and the dichotomy conditions on Mars can also be a reason for the dust devils' formations in the huge amount.

## 6. Discussion

Our study try to show that the dust devil's locations on Mars not just random anymore because the necessary conditions could be the intense surface heating due to sun and also the hemispherical position of Mars during the different seasons for the formation but the sufficient conditions may contain some more factors like the geographical conditions including the Mars remanent magnetic fields and the unknown conditions on nearby regions of dichotomy of Mars so, we try to show and demonstrates a correlation between the locations of dust devil occurrences and the crustal magnetic fields zones and a pattern that followed by the DDs which is like the serpent like shape of the dichotomy that sharp boundary line which is followed by the DDs throughout the Mars. Our data shows that There are ~76.25% of Dust Devils are found on Southern Hemisphere of Mars. (Including Direct and Indirect Observations) and there are ~75.00% of Active Dust Devils are found on southern Hemisphere of Mars. (Including Direct and Indirect Observations which contains observations of Active Dust Devils means they are captured during their activity on Mars).

Our results show this previously unknown behavior of these dust devils on, because these dust devils' activity is quite frequent and highly active during the summer seasons in the southern hemisphere of Mars, but the pattern of occurrence mostly followed by these dust devils throughout the dichotomy and most probably found on the magnetic field zones which may have the combined effects on these dust devils.

Our study is limited to the observations and as we are developing more generalized way to including more & more observations to get to know why these dust devils behaving in a way the way it is and does there are any other hidden patterns which these dust devils are following.

## 7. Conclusion

The results that we have obtained from our observations & survey shows that the most of the dust devil's activity locations are related to the locations of crustal magnetic fields zones and follows the dichotomy of Mars and as we increase the data samples of these dust devils' activity locations, we get more and more accuracy in our observations and get to know the more hidden patterns that these dust devil's may be following.

And this relation shows a new possibility to look forward the dust devils and their formation on certain locations on Mars may not only just because of surface heating and the appropriate conditions for the formation of dust devils on but also due to some other factors like crustal magnetic fields and the dichotomy conditions in the regions also plays an important role.

## References

Watters, T. R., McGovern, P. J., & Irwin III, R. P. (2007). Hemispheres Apart: The Crustal Dichotomy on Mars. Annual Review of Earth and Planetary Sciences, 35(1), 621–652. https://doi:10.1146/annurev.earth.35.031306.140220

Carr MH. 1981. The Surface of Mars. New Haven, CT: Yale Univ. Press. 232 pp.

Mutch TA, Arvidson RE, Head JW, Jones KL, Saunders RS. 1976. The Geology of Mars. Princeton, NJ: Princeton Univ. Press

Balme, M., & Greeley, R. (2006). Dust devils on Earth and Mars. Reviews of Geophysics, 44, RG3003. https://doi.org/10.1029/2005RG000188

Kanak, K. M. (2005). Numerical simulation of dust devil-scale vortices. Quarterly Journal of the Royal Meteorological Society, 131, 1271–1292. https://doi.org/10.1256/qj.03.172

Spiga, A., Barth, E., Gu, Z., Hoffmann, F., Ito, J., Jemmett-Smith, B., et al. (2016). Large-Eddy Simulations of Dust Devils and Convective

Vortices. Space Science Reviews, 203(1-4), 245–275. https://doi.org/10.1007/s11214-016-0284-x

Toigo, A. D., Richardson, M. I., Ewald, S. P., & Gierasch, P. J. (2003). Numerical simulation of Martian dust devils. Journal of Geophysical Research, 108(E6), 5047. https://doi.org/10.1029/2002JE002002

Newman, C. E., Kahanpää, H., Richardson, M. I., Martínez, G. M., Vicente-Retortillo, A., & Lemmon, M. T. (2019). MarsWRF Convective Vortex and Dust Devil Predictions for Gale Crater Over 3 Mars Years and Comparison With MSL-REMS Observations. Journal of Geophysical Research: Planets, 124(12), 3442–3468. Portico. https://doi.org/10.1029/2019je006082

B. R. W. J. B. P. J. D. I. R. N. L. R. Greeley, "Dust Storms on Mars: Consideration and Simulations," NASA Technical Memorandum 78423, 1997

Perrin, C., Rodriguez, S., Jacob, A., Lucas, A., Spiga, A., Murdoch, N., … Banerdt, W. B. (2020). Monitoring of Dust Devil Tracks Around the InSight Landing Site, Mars, and Comparison with in-situ Atmospheric Data. Geophysical Research Letters. https://doi:10.1029/2020gl087234

Connerney, J. E. P., Acuna, M. H., Ness, N. F., Kletetschka, G., Mitchell, D. L., Lin, R. P., & Reme, H. (2005). Tectonic implications of Mars crustal magnetism. Proceedings of the National Academy of Sciences, 102(42), 14970–14975. https://doi:10.1073/pnas.0507469102

Acuña, M. H. (2003). The magnetic field of Mars. The Leading Edge, 22, 769–771.

Acuña, M. H., Connerney, J. E. P., Ness, N. F., Lin, R. P., Mitchell, D., Carlson, C. W., et al. (1999). Global distribution of crustal magnetization discovered by the Mars Global Surveyor MAG/ER experiment. Science, 284, 790. https://doi.org/10.1126/science.284.5415.790

Acuña, M. H., Connerney, J. E. P., Wasilewski, P., Lin, R. P., Anderson, K. A., Carlson, C. W., et al. (1998). Magnetic field and plasma observations at Mars: Initial results ofMars Global Surveyor mission. Science, 279, 1676–1680. https://doi.org/10.1126/science.279.5357.1676

Dolginov, S. S., & Zhuzgov, L. N. (1991). The magnetic field and magnetosphere of the planet Mars. Planetary and Space Science,39, 1493–1510.

Moehlmann, D., Riedler, W., Rustenbach, J., Schwingenschuh, K., Kurths, J., Motschmann, U., et al. (1991). The question of an internal Martian magnetic field. Planetary and Space Science, 39, 83–88.

Slavin, J. A., Schwingenschuh, K., Riedler, W., & Eroshenko, E. (1991). The solar wind interaction with Mars-Mariner 4, Mars 2, Mars 3, Mars 5, and PHOBOS 2 observations of bow shock position and shape. Journal of Geophysical Research, 96, 11,235–11,241.

Moehlmann, D. (1992). The question of a Martian planetary magnetic field. Advances in Space Research, 12, 213–217.

Langlais, B., Thébault, E., Houliez, A., Purucker, M. E., & Lillis, R. J. (2019). A new model of the crustal magnetic field of Mars using MGS and MAVEN. Journal of Geophysical Research:Planets. https://doi:10.1029/2018je005854

Hepburn, A. J., Holt, T., Hubbard, B., & Ng, F. (2019). Creating HiRISE digital elevation models for Mars using the open-source Ames Stereo Pipeline. Geoscientific Instrumentation, Methods and Data Systems Discussions, 1–31. https://doi:10.5194/gi-2019-11

Reiss, D., Fenton, L., Neakrase, L., Zimmerman, M., Statella, T., Whelley, P., … Balme, M. (2016). Dust Devil Tracks. Space Science Reviews, 203(1-4), 143–181. https://doi:10.1007/s11214-016-0308-6

Smith, D. E., Zuber, M. T., Solomon, S. C., Phillips, R. J., Head, J. W., Garvin, J. B., Banerdt, W. B., Muhleman, D. O., Pettengill, G. H., Neumann, G. A., et al. (1999) Science 284, 1495–1503.

USGS Astrogeology Science Center, 2009, Mars Viking Colorized Global Mosaic 232m v2 | USGS Astrogeology Science Center, NASA AMES , accessed 01 Sep 2023, https://astrogeology.usgs.gov/search/map/Mars/Viking/MDIM21/Mars_Viking_MDIM21_ClrMosaic_global_232m

U.S. Geological Survey, 2000, Mars Orbiter Laser Altimiter (MOLA), Mercator map of Mars without polar regions, NASA/JPL/GSFC, accessed 01 Sep 2023, https://commons.wikimedia.org/wiki/File:Mars_topography_(MOLA_dataset)_HiRes.jpg